\documentclass[a4paper,11pt]{article}
\pdfoutput=1 

\usepackage{jinstpub} 

\newcommand{\up}[1]{\textcolor{black}{#1}}

\title{\boldmath  Performances of two resistive MicroMegas prototypes for the Time Projection Chambers of the T2K Near Detector upgrade}

\author[1] {C.Jes\'{u}s-Valls,\note{Corresponding author.}}

\affiliation[1]{Institut de F\'{i}sica d'Altes Energies (IFAE), Carrer Can Magrans s/n, Edifici Cn, UAB campus, E-08193 Bellaterra (Barcelona), Spain.}

\emailAdd{cesar.jesus@cern.ch}

\abstract{T2K is a long baseline neutrino experiment that has been operating since 2009 providing some of the world-wide leading measurements for neutrino oscillation parameters. An upgrade for the Near Detector, ND280, of T2K has been proposed. It includes the installation of two new Time Projection Chambers (TPC) based on a new read-out technology: the resistive anode MicroMegas (RMM). This \up{technology} is expected to reduce the number of electronic channels while maintaining or improving the current ND280 TPCs' bulk MicroMegas perfomance. A series of tests have been \up{performed} and further \up{studies} are ongoing to validate this approach. \up{Two RMM prototypes were studied in dedicated beamtests. All results so far, some of them still preliminary, indicate that a better performance can be achieved, even with larger pads, thanks to the advantages of RMM.}}


\collaboration[c]{on behalf of the ND280 upgrade HA-TPC working group}

\proceeding{N$^{\text{th}}$ Instrumentation for Colliding Beam Physics (INSTR-20)\\
  24–28 February 2020\\
  Budker Institute of Nuclear Physics (BINP), Novosibirsk, Russia}

\begin{document}
\maketitle
\flushbottom

\section{Introduction}
T2K is a long baseline neutrino experiment located in Japan ~\cite{t2k_exp}. An upgrade project is ongoing towards T2K-II including a beam power increase ~\cite{t2k_upbeam_conf,t2k_upbeam_tdr} and an upgrade of its near detector ND280 ~\cite{nd280up_cdr,nd280up_tdr}. 
The modifications will reduce the systematic errors from the current $\sim6\%$ to $\sim4\%$. 
This will allow to further determine the neutrino oscillation parameters and in particular to improve the measurements in $\delta_{\textup{CP}}$ for which T2K has recently reported matter-antimatter asymmetry indications excluding most values of $\delta_{\textup{CP}}$ in the interval [0,$\pi$] at 3$\sigma$~\cite{nature_dcp}. 
\section{The ND280 upgrade}
\begin{figure}[htbp]
\centering 
\includegraphics[width=.4\textwidth,origin=c,angle=0]{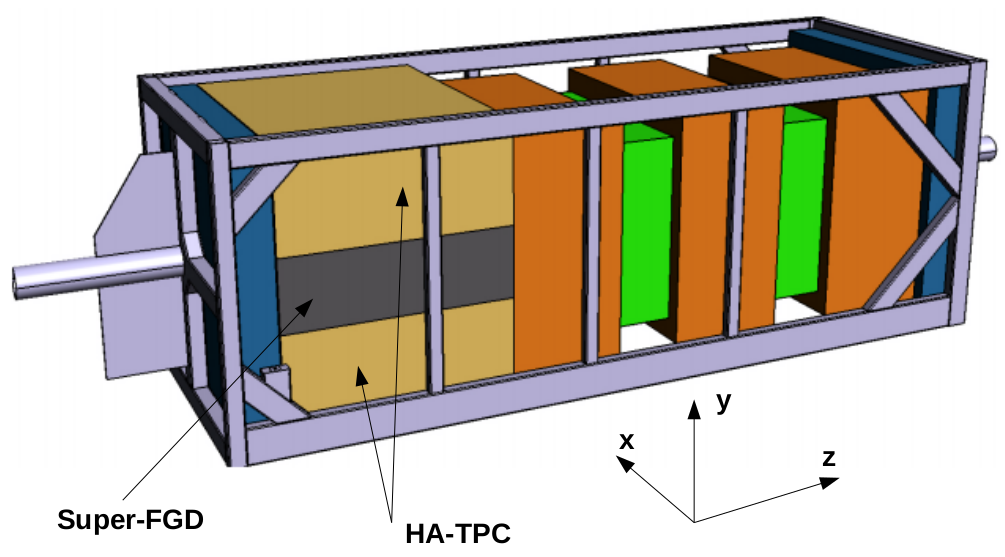}
\qquad
\includegraphics[width=.4\textwidth,origin=c,angle=0]{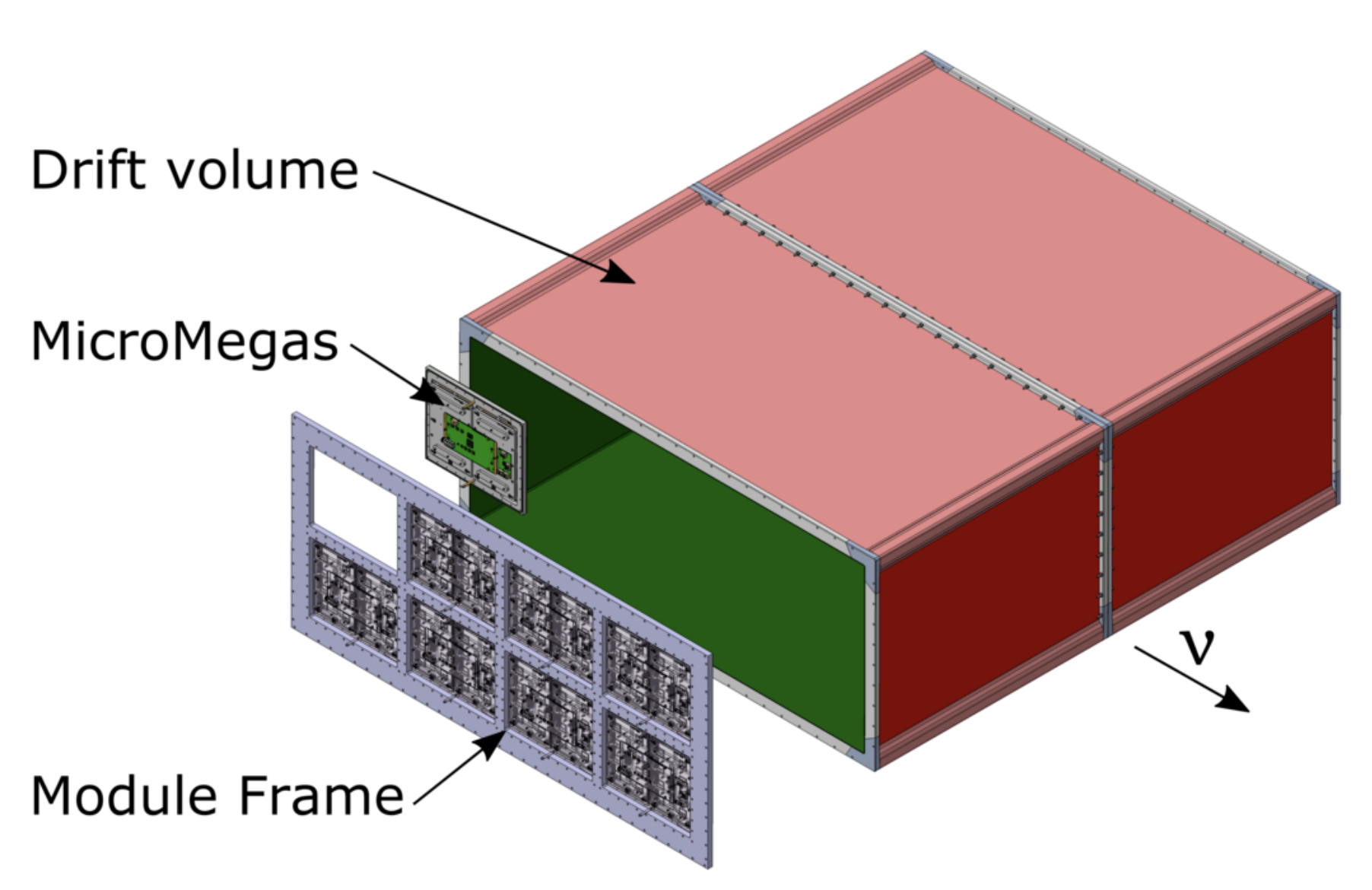}
\caption{\label{fig:nd280upgrade} {\bf Left}: Sketch of the upgrade ND280 basket without the ToF panels. The geometry of the new tracking elements has been optimized to detect tracks at high angle w.r.t to beam direction (Z-axis). {\bf Right}: Sketch of one of the two identical HA-TPCs. The inner TPC volume is split in to halves by a central cathode defining two independent drift volumes. In each side a module frame holds a 2x4 array of resistive anode MicroMegas.}
\end{figure}

The current ND280 detector consists of a \up{$\pi^0$} detector (P0D) ~\cite{nd280_p0d} which is based on water and scintillator materials with the main purpose to measure neutral current events and a tracker complex with two fine grained detectors (FGD) ~\cite{nd280_fgd} and three time projection chambers (TPC) ~\cite{nd280_tpc} designed to measure the charge and momenta for leptons and hadrons produced in charged current neutrino interactions. The P0D and the tracker are surrounded by an electromagnetic calorimeter (ECAL) ~\cite{nd280_ecal} which is enclosed in the former UA1/NOMAD dipole magnet. In addition, Side Muon Range Detectors (SMRD) ~\cite{nd280_smrd} are embedded in the iron yokes of the magnet. For the ND280 upgrade the P0D will be replaced by a new tracker system consisting of a Super Fine Grained Detector (SuperFGD), two horizontal TPCs (HA-TPC) and six Time-of-Flight (ToF) panels.\up{A detailed explanation about the ND280 upgrade and all its new modules (SuperFGD,HA-TPCs and ToF) can be found at~\cite{nd280up_tdr}}. A sketch of the upgrade configuration of all elements inside the ECAL is presented in the left hand side of Figure~\ref{fig:nd280upgrade}.
The upgraded detector will have twice the target mass of the current FGDs, will improve the detection efficiency for leptons emitted at large angles with respect to the neutrino beam direction and will allow for a better reconstruction of the hadronic part of the interaction, thanks to the \up{ 3D fine granularity} of the SuperFGD~\cite{nd280up_tdr,neutrons_sfgd,sfgd_proto}.
\subsection{The new HA-TPCs}
In order to identify the tracks exiting SuperFGD at high angle two new TPCs will be installed. The new high angle TPCs (HA-TPCs) will have a central cathode, a maximum drift distance of \up{90~cm} and resistive Micromegas (RMM) read-out on both sides. The HA-TPCs field cage~\cite{NP07} will be made of a thin, \up{4cm}, multi-layer composite material with low, $4\%$, radiation length to minimize the role of the dead material between the target and the TPCs. Each TPC anode will hold 8 RMM in a 2x4 array, such that 32 RMM will be installed in total. It is expected for them to provide good spatial and dE/dx resolution, similar or better than the current ND280 TPCs~\cite{nd280_tpc}, to accurately determine both the track curvature and its ionization to perform particle identification. The HA-TPCs design is presented in the right hand side of Figure~\ref{fig:nd280upgrade}.
\section{A resistive anode MicroMegas read-out}
The \textit{bulk} technology for building MicroMegas detectors has been used in ND280 since 2009~\cite{bulk_MM}. An improvement with respect to this technology, the \textit{resistive anode} MicroMegas (RMM), can achieve higher spatial resolution by means of charge sharing among the pads~\cite{RMM}. In the case of T2K, the resolution in reconstructing neutrino energy is limited by the Fermi motion of nuclei inside the nucleus, thus reducing the benefits of an improved spatial resolution. Nevertheless, RMM for T2K will allow to keep the original spatial resolution capabilities by using fewer and larger pads, thus reducing the number of electronics channels. So far, in the process of developing the final RMM model for the new ND280 HA-TPCs two prototype designs have been built and test, called MM0 and MM1.
\up{
\subsection{The RMM working principle}
When a track crosses the gas volume of a TPC it generates a cloud of ionized electrons. This electrons are driven to the anode using a drift field and under ideal gas conditions as much electrons as originally generated reach the anode. In the anode, the MicroMegas reads the signal. It consists of a mesh and a set of pads, in-between of which there is a higher electric field that generates an electron avalanche. For the bulk MicroMegas the electon avalanche is quite narrow w.r.t the pad size and therefore the position resolution is often limited by the pad size. In the resistive MicroMegas, the avalanche is spread over an insulator according to the telegraph equation~\cite{RMM}
\begin{equation}
\label{eq:RMM}
\rho(x,t) = \frac{1}{{\sigma^2_r}} e^{\frac{-r^2}{2{\sigma^2_r}}},\qquad \sigma_r = \sqrt{\frac{2t}{RC}}
\end{equation}
inducing a signal in several pads providing a more complete information about the track position. If the spread $\sigma_r$ is larger the signal reaches pads further away. The sketch of the bulk and the resistive MicroMegas concepts is presented in Figure~\ref{fig:MM_sketches}. 
\begin{figure}[htbp]
\centering
\includegraphics[width=.8\textwidth,origin=c,angle=0]{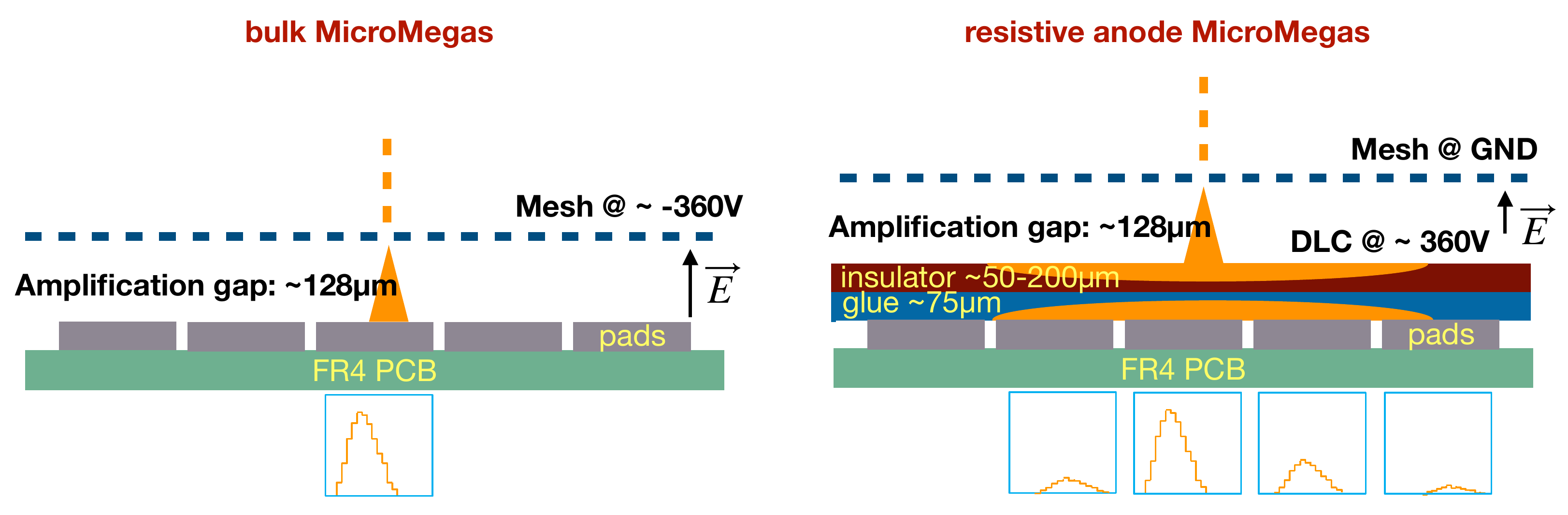}
\caption{\label{fig:MM_sketches} The specifications match those in the current ND280 MicroMegas and in the tested prototypes.  {\bf Left}: Standard ND280 TPC Bulk MicroMegas layout. {\bf Right}: ND280 upgrade HA-TPCs resistive anode MicroMegas layout. The insulating layer thickness determines the amount of spreading according to equation~\eqref{eq:RMM}.}
\end{figure}
}
\subsection{The MM0 prototype}
The MM0 prototype involved a standard T2K TPC bulk MicroMegas. This layout consists on a surface of 36$\times$34~cm$^2$ covered by 0.98$\times$0.70~cm$^2$. To make it resistive the pads were covered by a 200$\mu$m insulating layer acting as the capacitance, and then a 50~$\mu$m kapton (Apical) with a thin Diamond-LikeCarbon (DLC) layer providing a design resistivity of 2.5~M$\Omega/\square$. The electronics used were the same
that had been developed for the T2K TPCs based on the AFTER
chip [18]. The main goal of this prototype was to validate the RMM approach while developing and building the final layout. 
\subsection{The MM1 prototype}
The MM1 prototype corresponds to the final expected layout for the new ND280 TPCs read-out. It has a total surface of 34$\times$42~cm$^2$ organized in 32$\times$36~pads of 1.1$\times$1.0~cm$^2$. The pad surface is covered by a 75~$\mu$m insulating layer acting as the capacitance, and then a 50~$\mu$m kapton (Apical) with a thin DLC layer. The resistivity corresponds to the design value of 0.4~M$\Omega/\square$, hence providing larger sharing than MM0. The current ND280 TPCs' electronics were adapted to the new layout based on the same chip (AFTER), as explained in~\cite{nd280up_tdr}.
\section{The MM0 beam test}
In summer 2018 the MM0 prototype was mounted in the HARP TPC~\cite{harp_tpc} an exposed to a beam of charged particles in the T9 area at CERN, see the full discussion at~\cite{btest_RMM}. This beam consisted on a mixture of pions, muons, electrons and protons, and a trigger system was prepared to classify the different events. The data was taken without a magnet and a  $^{55}$Fe source was placed in the cathode to monitor the gain. In addition, a cosmic trigger was settled. As explained in ~\cite{btest_RMM} the gas quality smoothly decreased along the data taking due to the accumulation of impurities in the gas. This effect was corrected in the analysis by computing correction factors extracted from gain and cosmic data.
\subsection{Gain and uniformity}
\begin{figure}[htbp]
\centering 
\includegraphics[width=.49\textwidth,origin=c,angle=0]{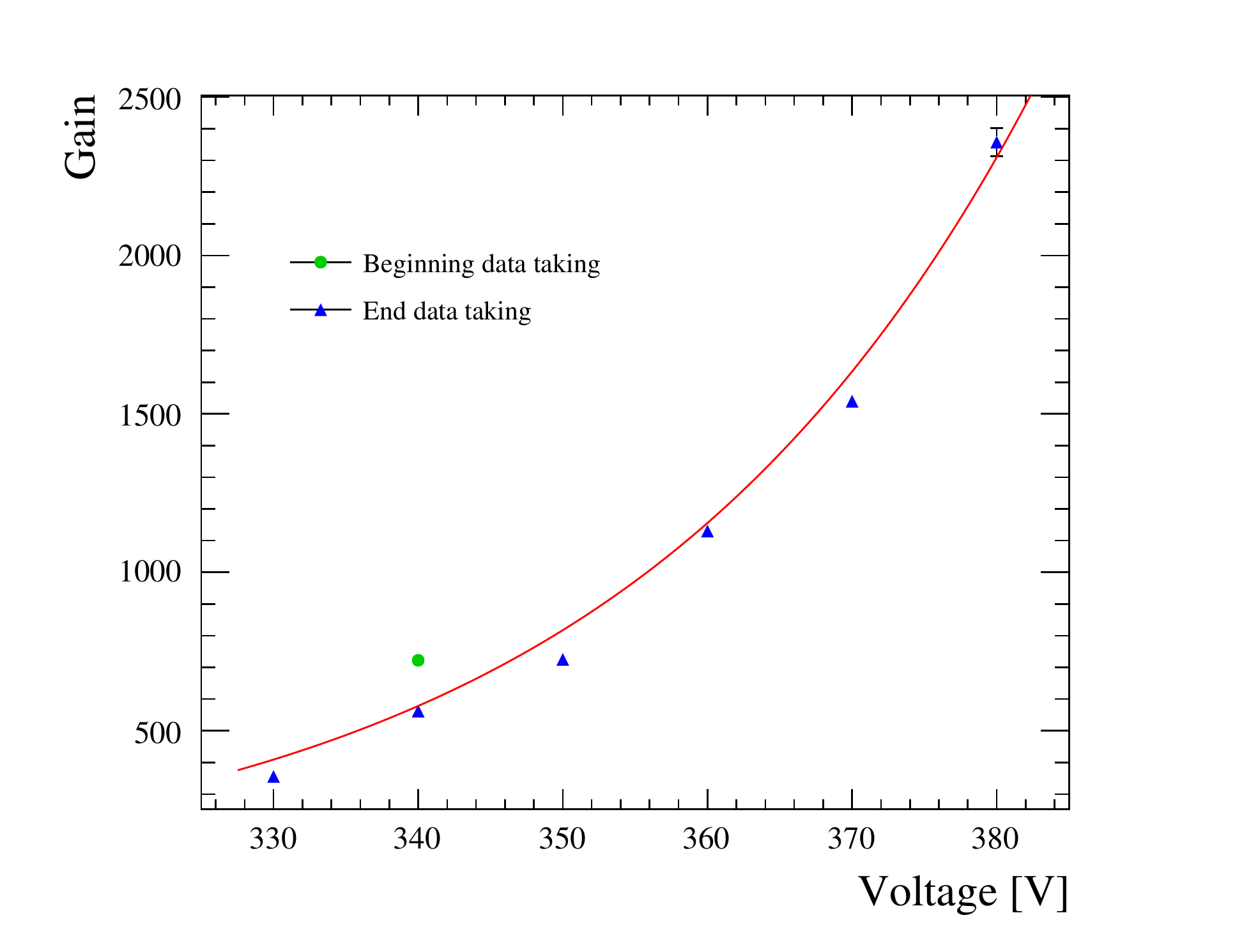}
\includegraphics[width=.49\textwidth,origin=c,angle=0]{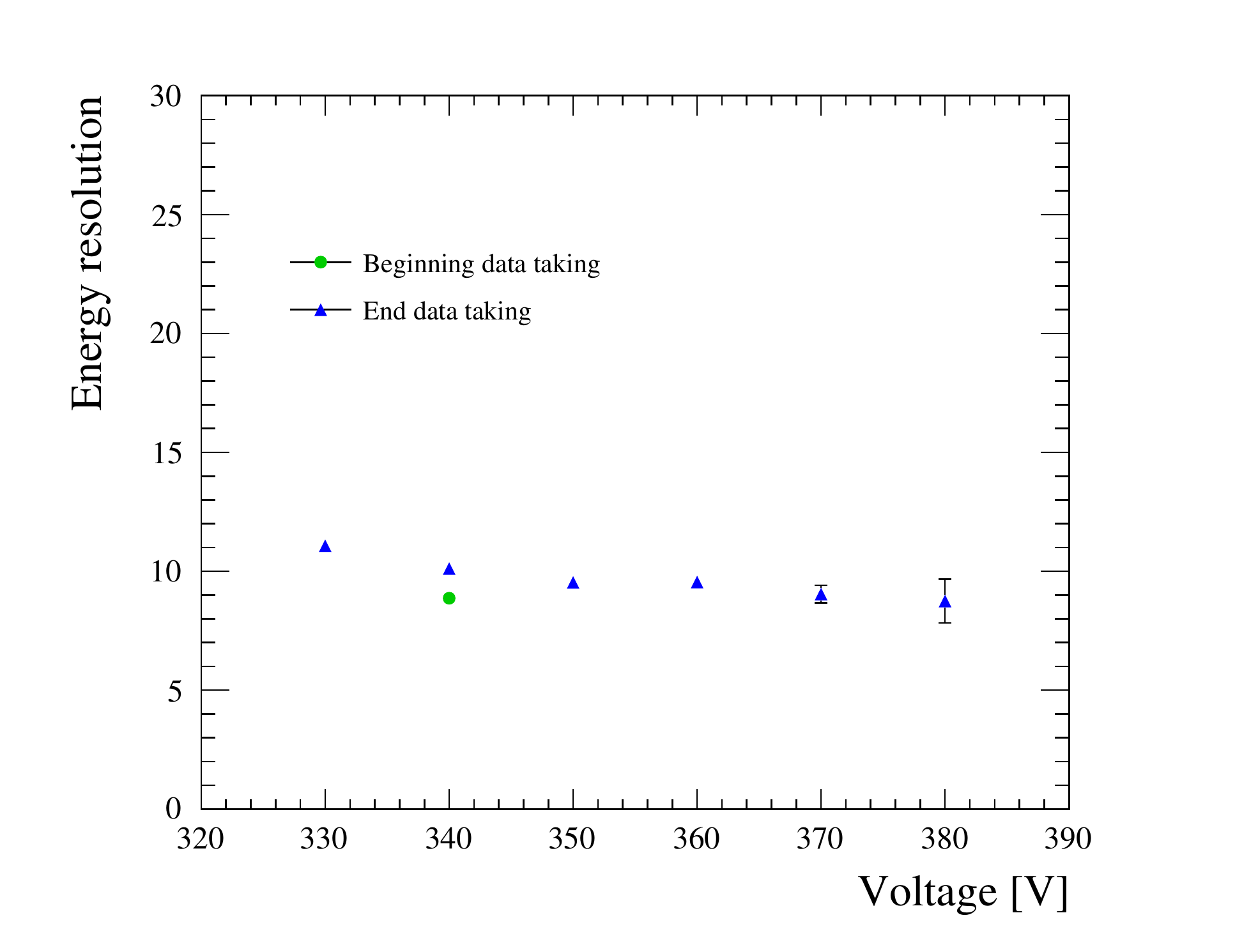}
\caption{\label{fig:CERN_RMM_gain} {\bf Left}: Absolute Gain dependence with the mesh Voltage. {\bf Right}: Energy resolution of the $^{55}Fe$ 5.9~keV signal. The data at the beginning of the data taking corresponds to the best gas conditions.}
\end{figure}

The MM0 measured gain performance is shown in Figure~\ref{fig:CERN_RMM_gain}. The most important remarks are:
\begin{itemize}
    \item The $^{55}$Fe 5.9~keV signal was measured with 8.9$\%$ resolution ($\sigma/\mu$) even under non optimal gas conditions. This value is similar to the $8\%$ value measured in current the ND280 bulk MicroMegas.
    \item The absolute gain increased exponentially with the mesh Voltage. Stable operation without sparks was achived up to 380V.
    \item The channel uniformity, measured computing the average signal recorded with cosmic tracks was observed to be $3\%$, consistent with the current ND280 bulk MicroMegas.
\end{itemize}
\subsection{dE/dx}
\begin{figure}[htbp]
\centering
\includegraphics[width=.44\textwidth,origin=c,angle=0]{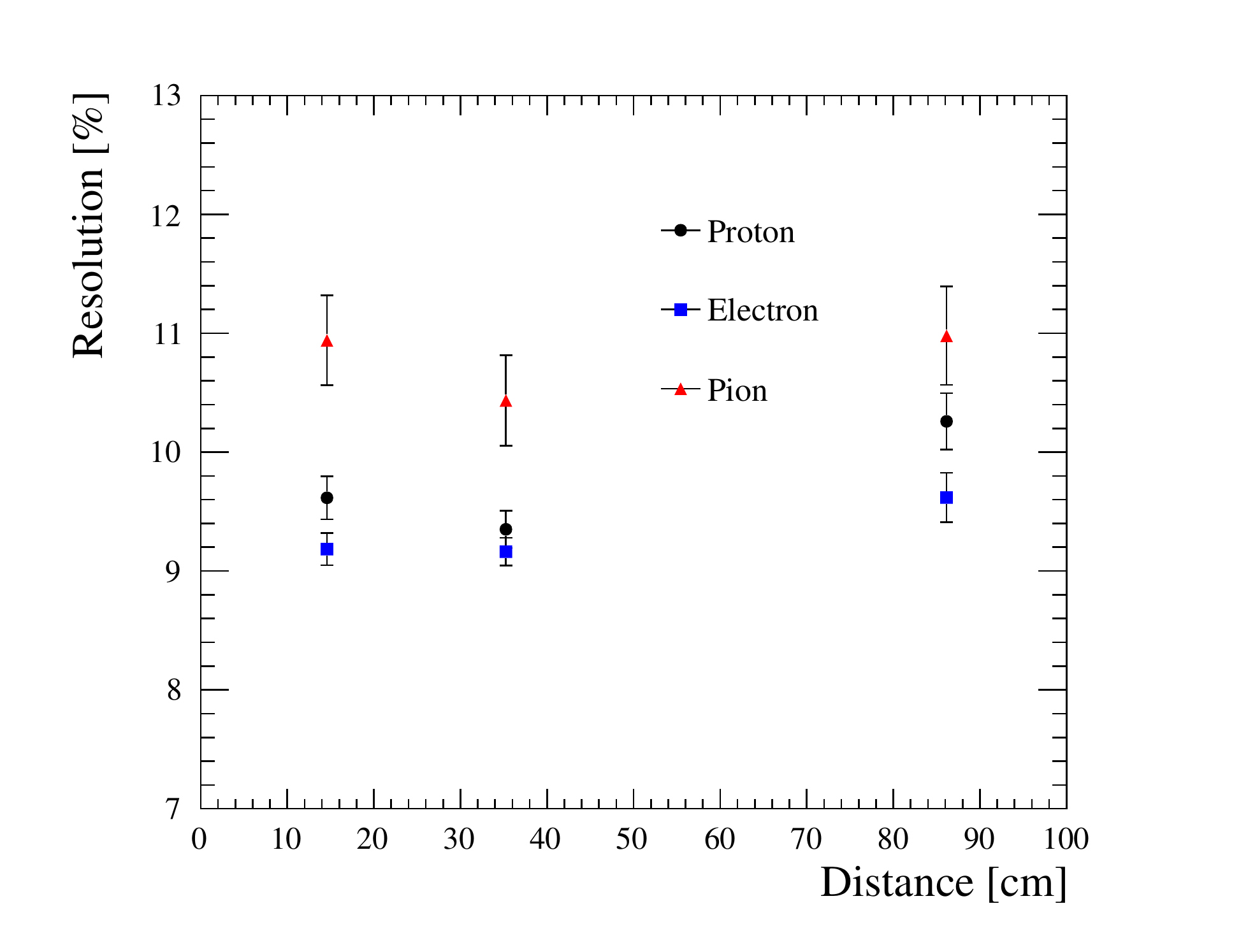}
\includegraphics[width=.55\textwidth,origin=c,angle=0]{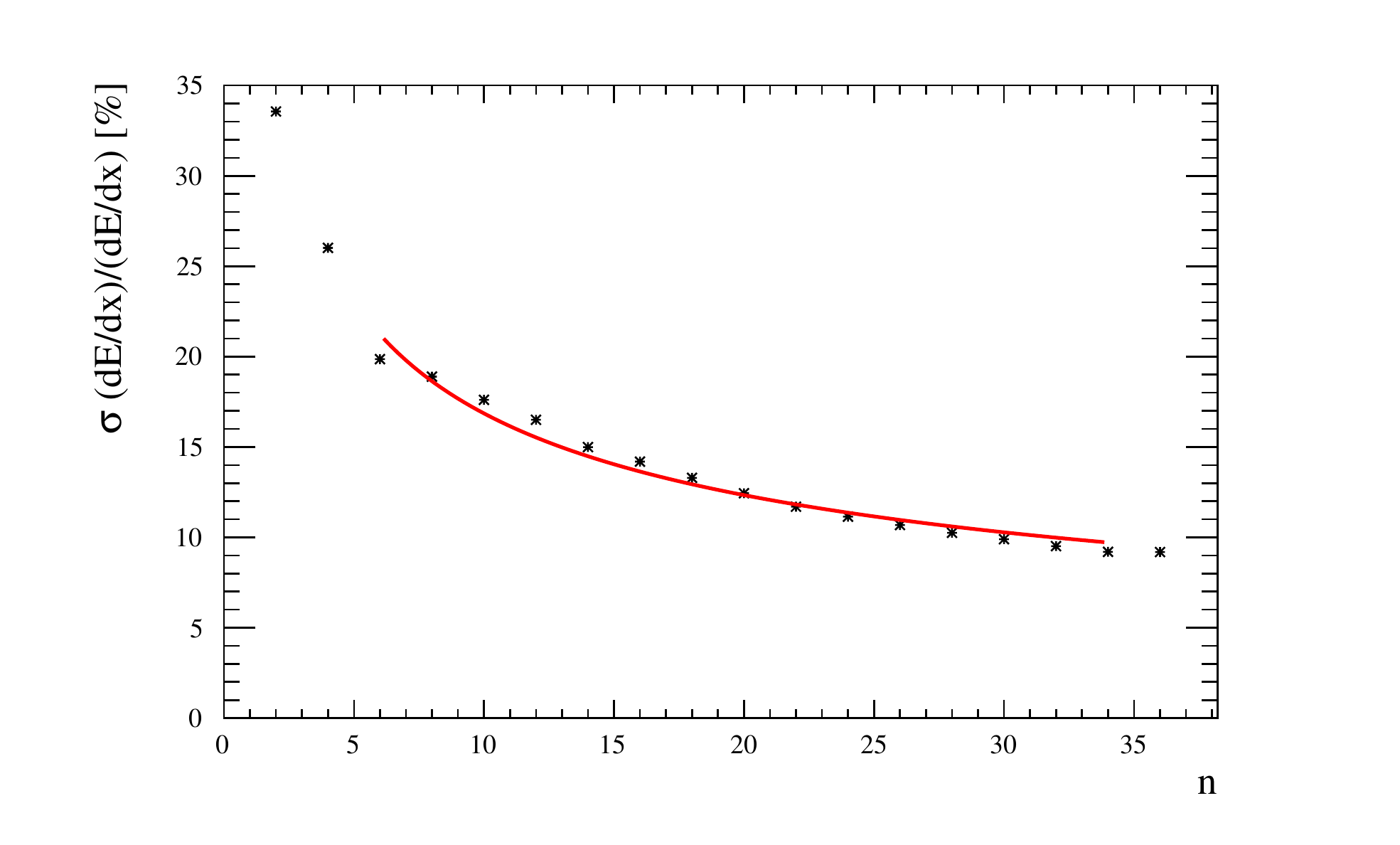}
\caption{\label{fig:CERN_dedx} {\bf Left}: dE/dx resolution ($\sigma/\mu$) for different particle types as a function of the drift distance. {\bf Right}:  dE/dx resolution ($\sigma/\mu$) as a function of the number of clusters used to compute it using the electron trigger and 0.8 GeV/c momentum.}
\end{figure}

The dE/dx for different particle triggers is shown in Figure~\ref{fig:CERN_dedx}. On it, the left hand side plot shows the stability of the dE/dx resolution ($\sigma/\mu$) versus the drift distance. The right hand side plot shows the dE/dx resolution as a function of the number of pad columns (a.k.a clusters) used to compute it. Using the fit, the extrapolated value using two MicroMegas (68-clusters) is estimated to be around $7\%$. In the current ND280 bulk MicroMegas, this value is $7.8\%$.
\subsection{Charge spreading}
\begin{figure}[htbp]
\centering
\includegraphics[width=.48\textwidth,origin=c,angle=0]{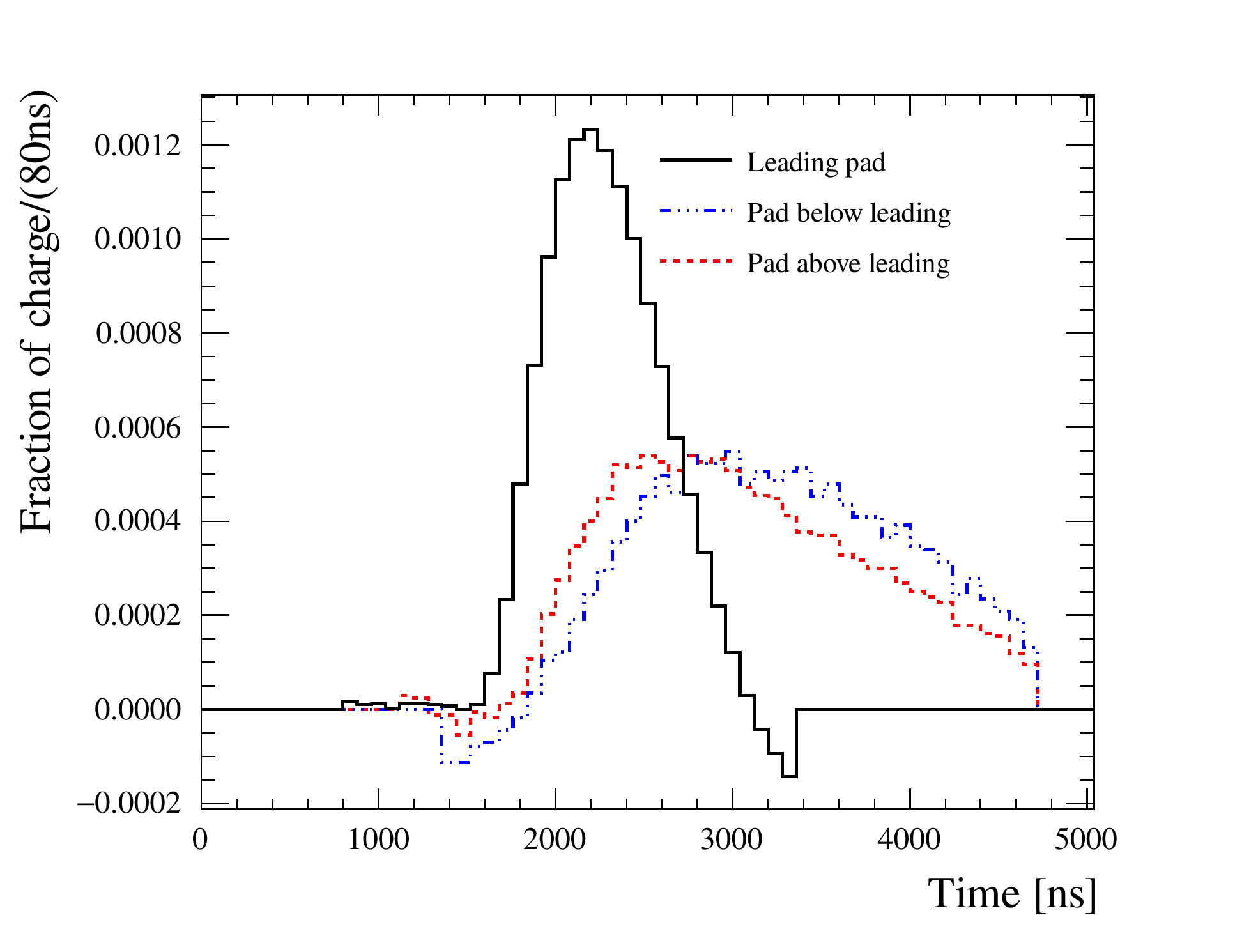}
\includegraphics[width=.48\textwidth,origin=c,angle=0]{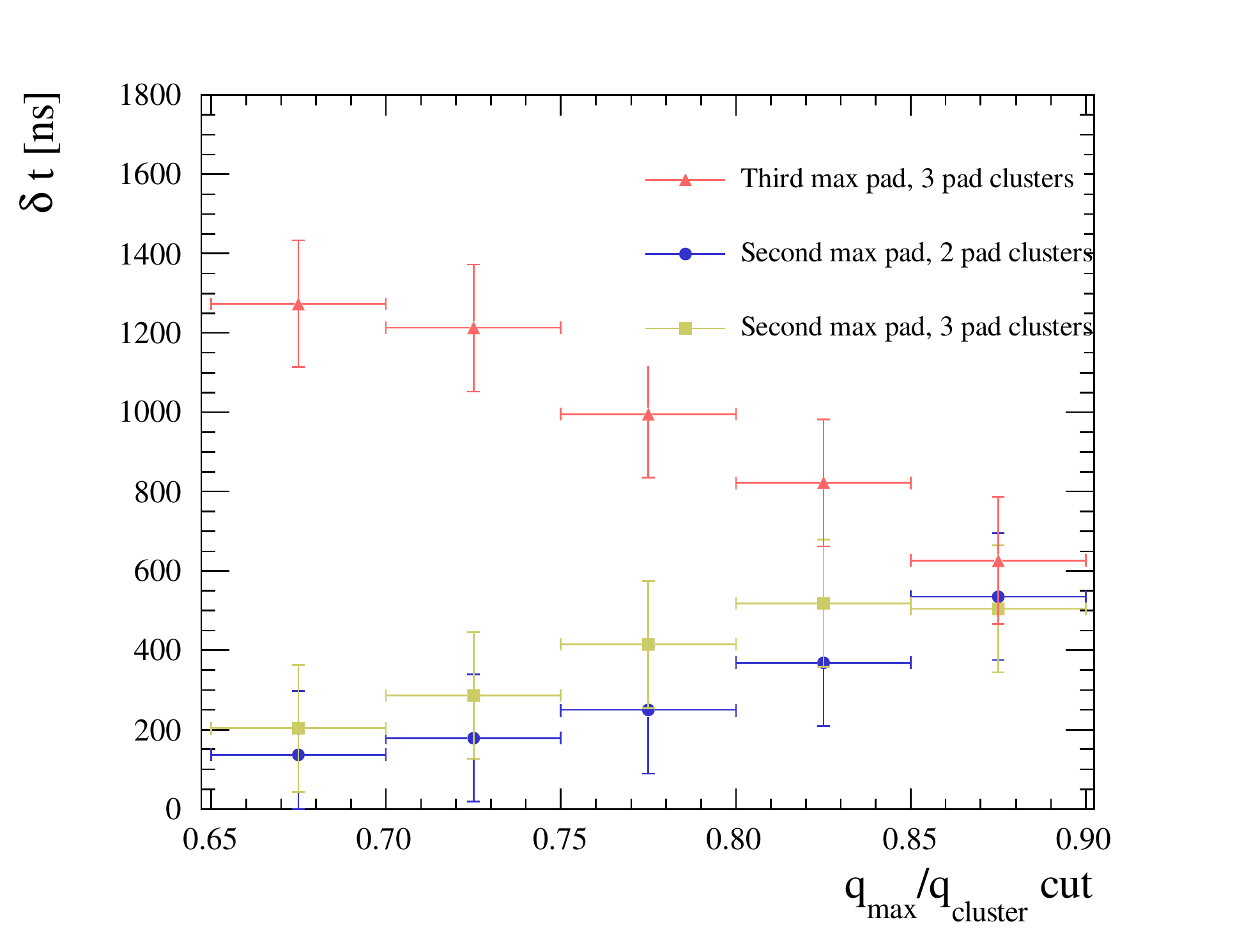}
\caption{\label{fig:CERN_ch_spread} {\bf Left}: Waveform example of a track signal over a 3-pad's cluster. The leading pad gets first most of the charge while the neighbor signals are lower and delayed to the extra charge spreading time. {\bf Right}: Three different measurements of the time delays between the leading pad and a neighbor pad. In the limit  $q_{\textup{max\_pad}}/q_{\textup{cluster}} \rightarrow 1$ all samples converge to $\delta_t$=600ns.}
\end{figure}

Measuring the charge spreading in a quantitative way is important to understand the charge sharing phenomena and to model the detector response in simulations. The charge spreading in the resistive anode generates signals in several pads. For a given cluster of column pads, the signal is first measured in the leading pad, i.e. the pad getting most of the charge, and collected latter by the pads surrounding it due to the extra time that the charge takes traveling from one pad to the other. If the track passes close to the leading pad's center most of the charge of the cluster is collected in it and the time delay in the neighbor pads is large, as presented in the example waveform in Figure~\ref{fig:CERN_ch_spread}. In contrast, if the track passes close to the edge of 2 pads the delay between the leading pad and the next-to leading pad is short, and the charge measured in both pads is very similar. In this way, the relation of the ratio $q_{\textup{max\_pad}}/q_{\textup{cluster}}$ and the associated time delay can be used to quantify the charge spread velocity $v_d$. This measurement is shown in the right hand side plot in Figure~\ref{fig:CERN_ch_spread}, where $v_d$ = 0.6~cm/$\mu$s is estimated by extrapolating the different samples time delay trend in the limit $q_{\textup{max\_pad}}/q_{\textup{cluster}} \rightarrow 1$, where the charge has to travel half the pad length.
\subsection{Spatial resolution}
\begin{figure}[htbp]
\centering
\includegraphics[width=.48\textwidth,origin=c,angle=0]{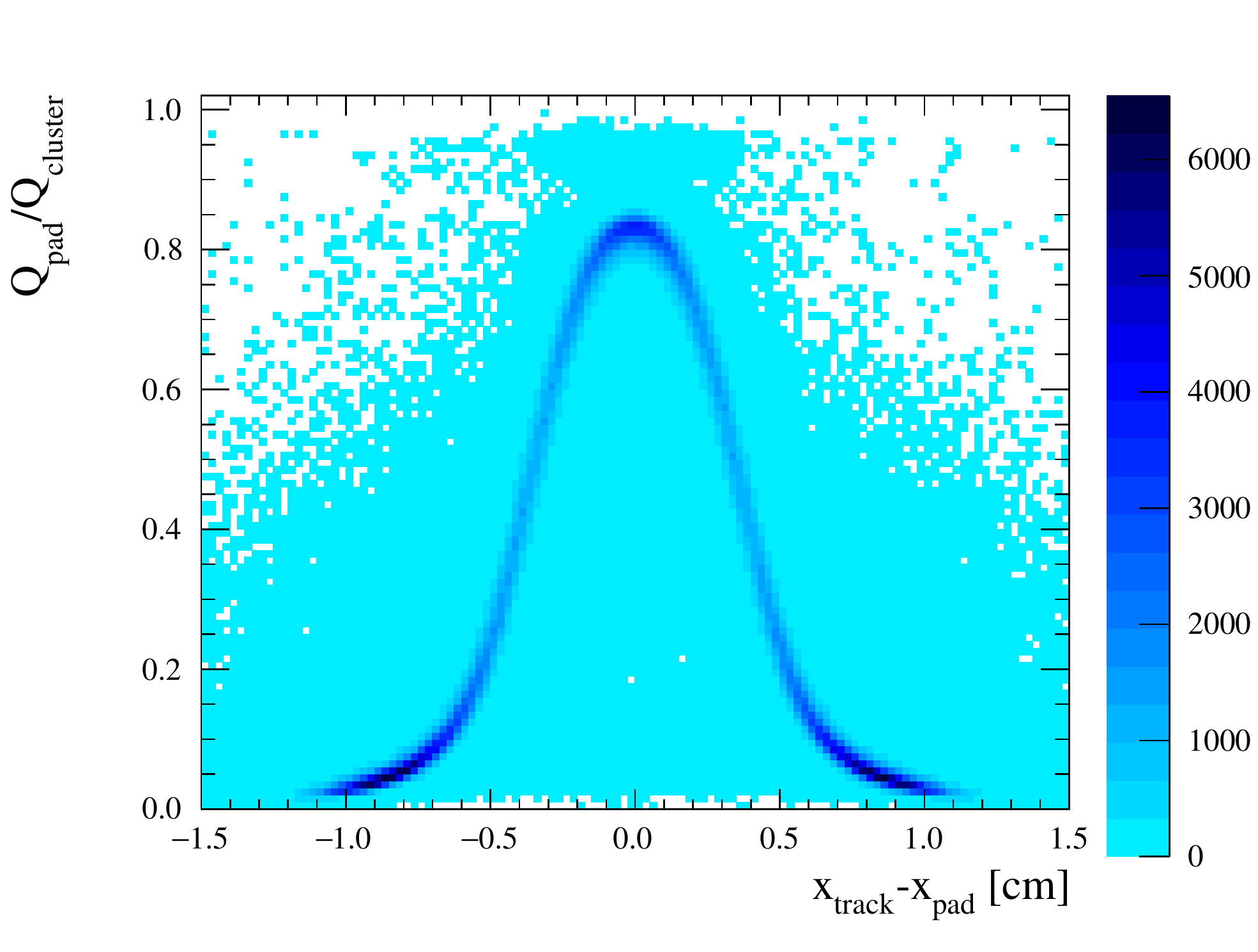}
\includegraphics[width=.48\textwidth,origin=c,angle=0]{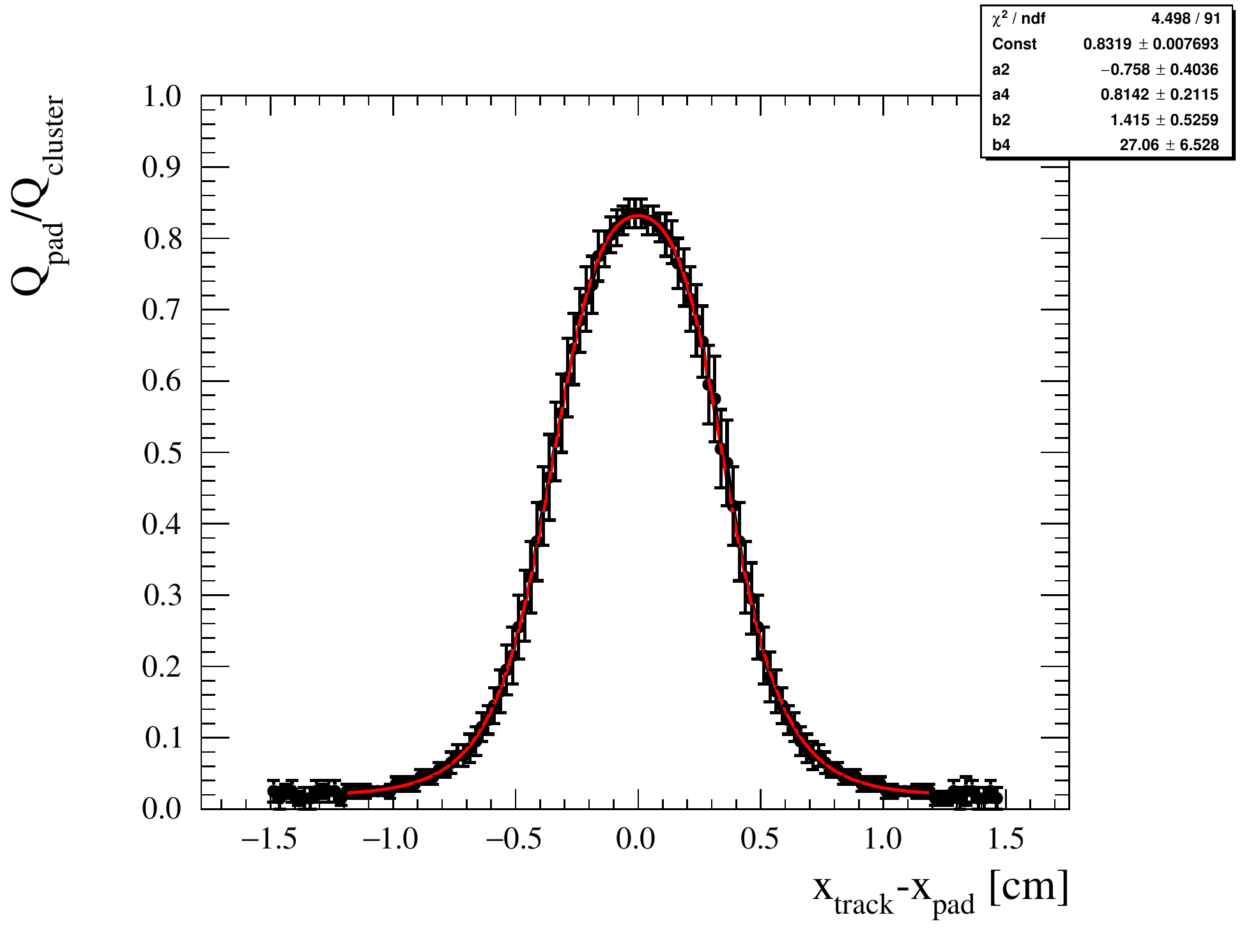}
\caption{\label{fig:CERN_PRF} {360V Mesh voltage, 1~GeV/c momentum, pion trigger, 30cm drift distance. {\bf Left:} 2D histrogram data. {\bf Right:} PRF with errors extracted from data and the analytical parametrization from equation~\eqref{eq:par}.}}
\end{figure}

\begin{figure}[htbp]
\centering
\includegraphics[width=.48\textwidth,origin=c,angle=0]{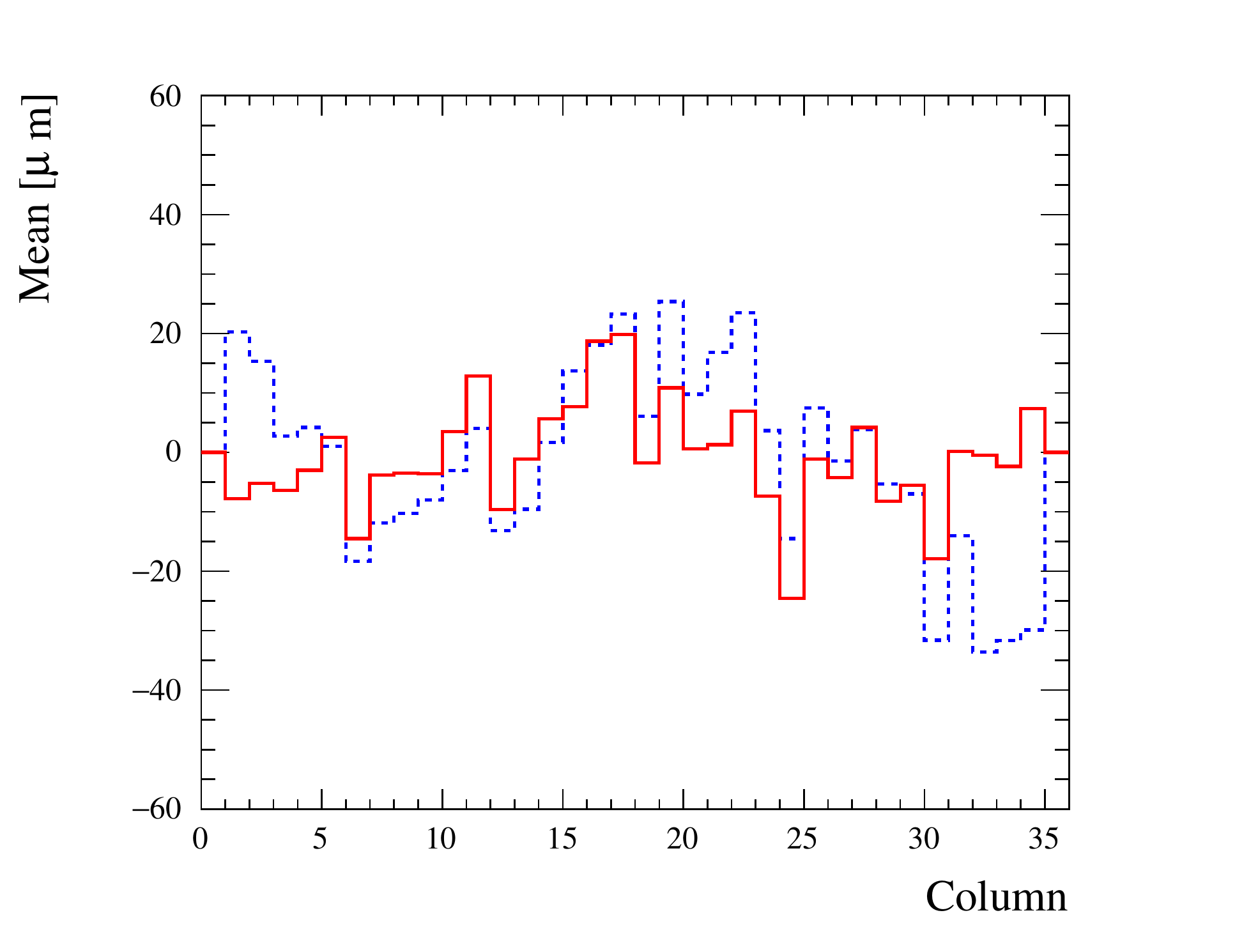}
\includegraphics[width=.48\textwidth,origin=c,angle=0]{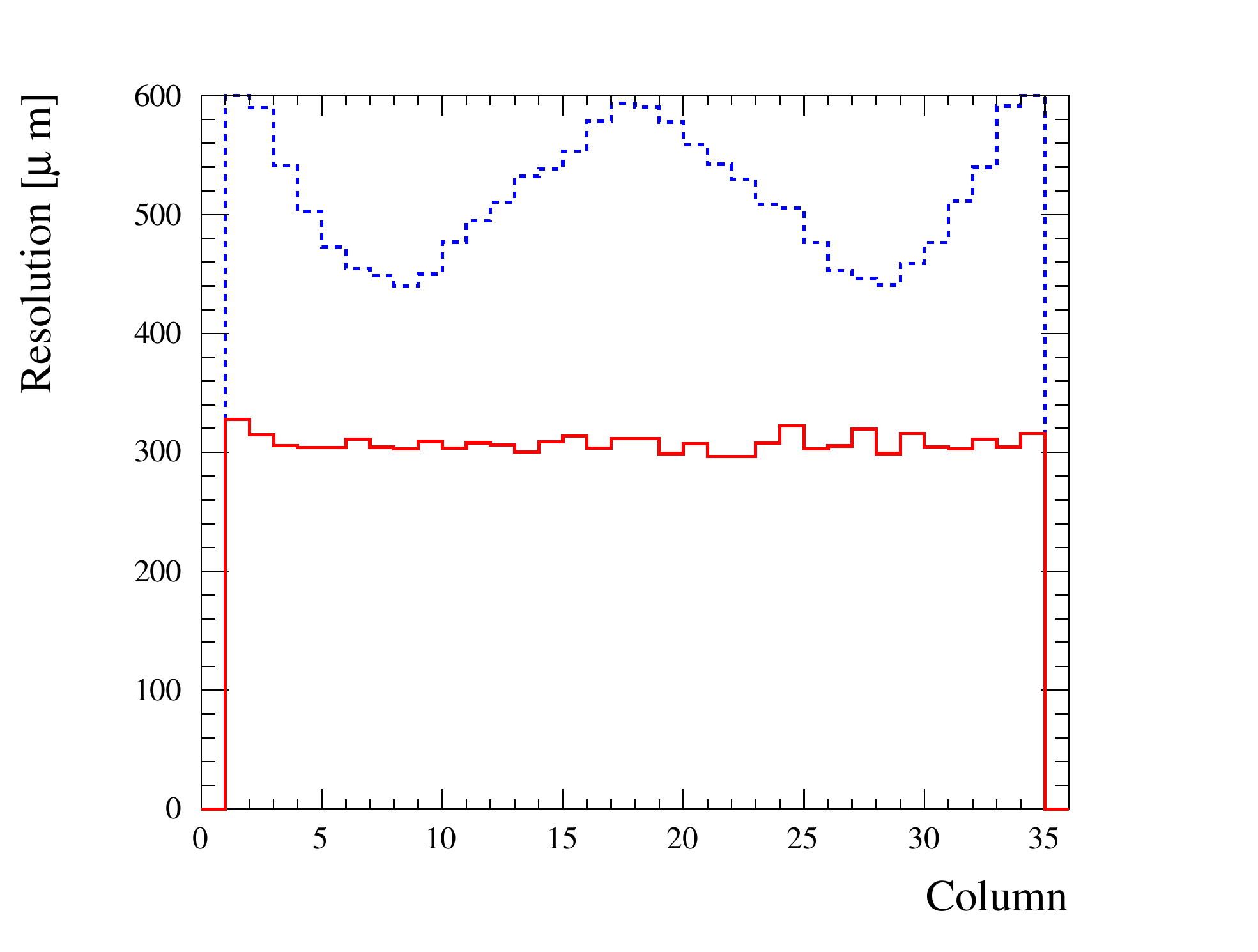}
\caption{\label{fig:CERN_spatial_res} 360V Mesh voltage, 1~GeV/c momentum, pion trigger, 30cm drift distance. {\bf Left:} Residuals mean. {\bf Right:} Spatial resolution.}
\end{figure}

The spatial resolution can be quantified by means of comparing the position estimate by a straight line fit using all clusters and the track position estimate in each individual cluster. The simplest estimate corresponds to the center of charge method (CoC) which computes the track position in each cluster weighting the pad's center position by its charge measurement. This estimator is however not the best one. The discretisation of a continuous distribution on a set of pads of finite size introduces a systematic bias. This, however, can be corrected by the use of a pad-response-function (PRF) relating the track position estimate $x_\textup{track}$, the pad center $x_\textup{pad}$ and the measured charge in each cluster's pad $Q_{\textup{pad}}$ w.r.t the total cluster's deposit $ Q_{\textup{cluster}}$,
\begin{equation}
    Q_{\textup{pad}}/ Q_{\textup{cluster}} = PRF (x_\textup{track}- x_\textup{pad}).
\end{equation}
The PRF is parametrized using the ratio of two symmetric 4th order polynoms proposed in [22]:
\begin{equation}
\label{eq:par}
PRF(x, \Gamma, \Delta, a, b)=\frac{1+a_2x^2+a_4x^4}{1+b_2x^2+b_4x^4}
\end{equation}
where the coefficients $a_2$,$a_4$, $b_2$ and $b_4$ can be expressed in terms of the full width half maximum $\Gamma$, the base width $\Delta$ of the PRF, and two scale parameters a and b.  The optimal parameters values are extracted minimizing the distance between the measurement and the PRF estimate:
\begin{equation}
\chi^2=\sum_{pads}\frac{Q_{pad}/Q_{cluster} - PRF\left(x_{track} - x_{pad}\right)}{\sqrt{Q_{pad}}/Q_{cluster}}.
\end{equation}
the PRF data and parametric function is shown in Figure~\ref{fig:CERN_PRF}. The residuals, $x_\textup{track}-x_\textup{pad}$, distribution is fit with a Gaussian from which the $\sigma$ defines the spatial resolution and the $\mu$ the position bias. The results using the CoC and the PRF methods are presented in Figure~\ref{fig:CERN_spatial_res} showing a stable 300~$\mu$m spatial resolution in all MM0 pad columns.
\section{The MM1 beam test}
The MM1 was test in a beam test at DESY during 2019's fall. The analysis of the collected data is still ongoing and it is expected to be fully presented in a future publication. In this beam test a magnet was equiped. The different runs collected data to analyze in detail the role of all the relevant parameters and it had the main goal to fully characterize the charge spreading, the resisitve foil uniformity and to ensure a performance satisfying the ND280 upgrade requirements with the increased pad size.
\subsection{MM1 preliminary performance results}
\begin{figure}[htbp]
\centering
\includegraphics[width=.48\textwidth,origin=c,angle=0]{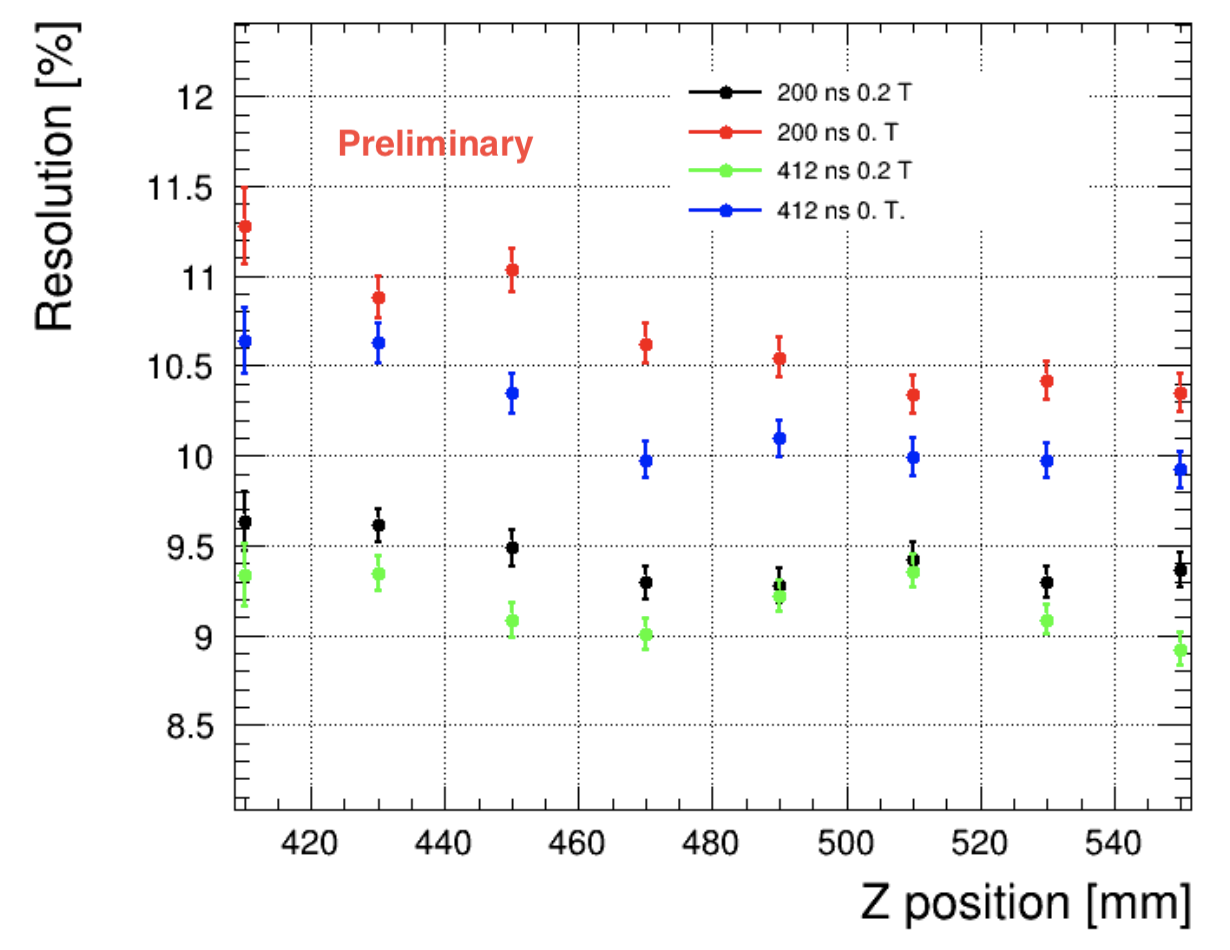}
\includegraphics[width=.48\textwidth,origin=c,angle=0]{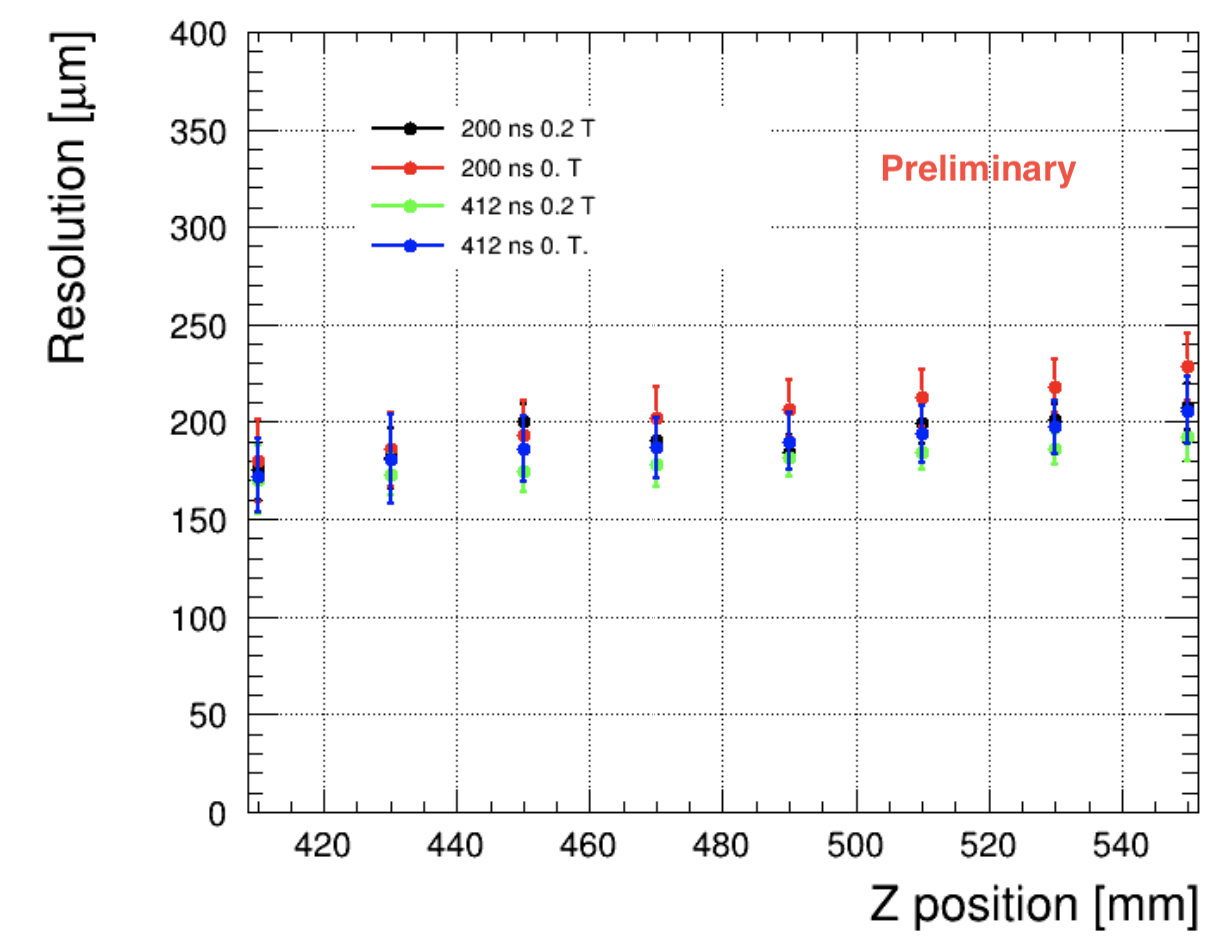}
\caption{\label{fig:DESY_preliminary} Drift distance, magnetic field and shaping time dependence. {\bf Left:} dE/dx resolution. {\bf Right:} Spatial resolution.}
\end{figure}

First preliminary results of the MM1 analysis are shown in Figure~\ref{fig:DESY_preliminary}. The modified layout with increased spreading shows an even better spatial resolution than the one observed in MM0 despite the increased pad size. The MM1 dE/dx resolution is similar to that measured with MM0, and shows a significant improvement under the presence of a 0.2T magnetic field, as expected. Both figures-of-merit well behave at different drift distances and are robust under shaping time choices.
\section{Conclusions}
The prototype developing and testing of resistive MicroMegas modules for the new HA-TPCs for the upgrade of the ND280 detector is ongoing. The MM0 results validated the resistive anode approach providing a satisfactory performance under the ND280 physics requirements~\cite{btest_RMM}. The first and preliminary results from the MM1 data analysis, including larger pad size, presents an even better performance with respect to the MM0 results due to the increased charge sharing. So far, all results support the viability of using resistive anode MicroMegas as a way to mantain an even improve the read-out performance while using less electronic channels.  The MM1 characterization using DESY data is expected to finish along 2020. A field cage prototype is currently being tested at CERN and a join test of the RMM read-out and the field cage prototypes is planed for 2020's fall at DESY. The final HA-TPC installation in Japan is scheduled \up{for} 2021.
\acknowledgments

This work has been founded by CEA and CNRS/IN2P3, France; DFG,
Germany; INFN, Italy; National Science Centre (NCN) and Ministry of Science and Higher Education, Poland; MINECO and ERDF funds, Spain. In addition, participation of individual researchers and institutions has been further supported by H2020 Grant No. RISE-GA644294-JENNIFER 2020. The measurements leading to these results have been performed at the Test Beam Facility at CERN Geneva (Switzerland), and at the Test Beam Facility at DESY Hamburg (Germany).


\end{document}